\documentclass[letter,traditabstract]{aa} 
\usepackage{graphicx}
\usepackage{txfonts}
\begin{document}
\titlerunning{Shedding new light on A1689 with {\em Herschel}}
\authorrunning{Haines et al.}
   \title{LoCuSS: Shedding New Light on the Massive Lensing Cluster Abell 1689 -- The View from {\em Herschel}\thanks{Herschel is an ESA space observatory with science instruments provided by European-led Principal Investigator consortia and with important participation from NASA.}}

   \author{C.~P.\ Haines\inst{1}, G.~P.\ Smith\inst{1},
     M.~J.\ Pereira\inst{2}, E. Egami\inst{2}, S.~M.\ Moran\inst{3},
     E.\ Hardegree-Ullman\inst{4}, T.\ D.\ Rawle\inst{2} 
     \and 
     M.\ Rex\inst{2}
          }

   \institute{School of Physics and Astronomy, University of Birmingham,
              Edgbaston, B15 2TT, UK. 
              \email{cph@star.sr.bham.ac.uk}
         \and     Steward Observatory, University of Arizona, 933 North Cherry Avenue, Tucson, AZ 85721, USA
         \and     Department of Physics and Astronomy, The Johns Hopkins University, 3400 N. Charles Street, Baltimore, MD 21218, USA
         \and
             Rensselaer Polytechnic Institute (RPI)  110 Eighth Street, Troy, NY 12180, USA\\
             }

   \date{Received March 28, 2010}

 
   \abstract {We present wide-field {\em Herschel}/PACS observations
     of A\,1689, a massive galaxy cluster at $z{=}0.1832$, from our
     Open Time Key Programme.  We detect 39 spectroscopically
     confirmed 100$\mu$m-selected cluster members down to
     $1.5{\times}10^{10}L_{\odot}$.  These galaxies are forming stars
     at rates in the range 1--10\,M$_{\odot}$/yr, and appear to
     comprise two distinct populations: two-thirds are unremarkable
     blue, late-type spirals found throughout the cluster; the
     remainder are dusty red sequence galaxies whose star formation is
     heavily obscured with A$({\rm H}\alpha){\sim}2$\,mag and are
     found only in the cluster outskirts. The specific-SFRs of these
     dusty red galaxies are lower than the blue late-types, suggesting
     that the former are in the process of being quenched, perhaps via
     pre-processing, the unobscured star formation being terminated
     first.  We also detect an excess of 100$\mu$m-selected galaxies
     extending ${\sim}6$Mpc in length along an axis that runs NE-SW
     through the cluster center at ${\ga}9$5\% confidence.  
 Qualitatively this structure is
     consistent with previous reports of substructure in X-ray,
     lensing, and near-infrared maps of this cluster, 
     further supporting the view that this cluster is a dynamically
     active, merging system.  }


   \maketitle
%

\section{Introduction}

Over the last 30 years a large body of evidence has formed that star
formation in cluster galaxies is much lower than in the surrounding field
(e.g. Lewis et al. \cite{lewis}).  A variety of mechanisms have been
proposed to quench star formation in infalling spirals such as
ram-pressure stripping or harassment (see Haines et
al. \cite{haines07} for a review), but the key distinguishing evidence
remains elusive.  However most of these studies rely on optical
spectroscopy using H$\alpha$ or the [O{\sc ii}] emission lines
(e.g. Poggianti et al. \cite{pog99}).  Recent mid-infrared based
studies with {\em ISO} and then {\em Spitzer} (e.g. Metcalfe et al. 
\cite{metcalfe}; Coia et al. \cite{coia}; Geach et al. \cite{geach06};   
Saintonge et al. \cite{saintonge}) have revealed a
significant population of dusty cluster galaxies whose
star formation is heavily obscured at optical wavelengths, leading to
[O{\sc ii}]-based SFRs underestimating the true level by
${\sim}1$0--3$0{\times}$ (Duc et al. \cite{duc}), while
the age-dependence of the dust obscuration can lead to dusty
star-bursts being in fact classified spectroscopically as
post-starburst galaxies (Poggianti \& Wu \cite{poggianti}).

Clusters are not isolated systems, but lie at the intersections of
filaments, constantly accreting galaxies and galaxy groups along these
filaments or directly from the field, and various studies suggest that
the key sites of galaxy transformation are within these infalling
structures rather than the cluster core (e.g.  Balogh et
al. \cite{balogh04}; Moran et al. \cite{moran}; Fadda et al. \cite{fadda08}).

These issues have motivated the Local Cluster Substructure
(LoCuSS\footnote{http://www.sr.bham.ac.uk/locuss}) \emph{Herschel} Key
Programme to study a large statistical sample of 30 massive galaxy
clusters at $z{\sim}0.2$ with panoramic FUV--FIR data from {\em GALEX}, {\em
  Spitzer} and {\em Herschel} (Haines et al. \cite{haines09a}; Smith
et al. \cite{smith10}).  These provide a complete census of star
formation in cluster galaxies extending to the infall regions, which
can then be related to the dynamical status of the cluster as determined
from complementary lensing, X-ray and dynamical analyses (Haines et
al. \cite{haines09b}; Pereira et al. \cite{pereira}).  

\begin{figure*}
\centering
\includegraphics*[height=101mm]{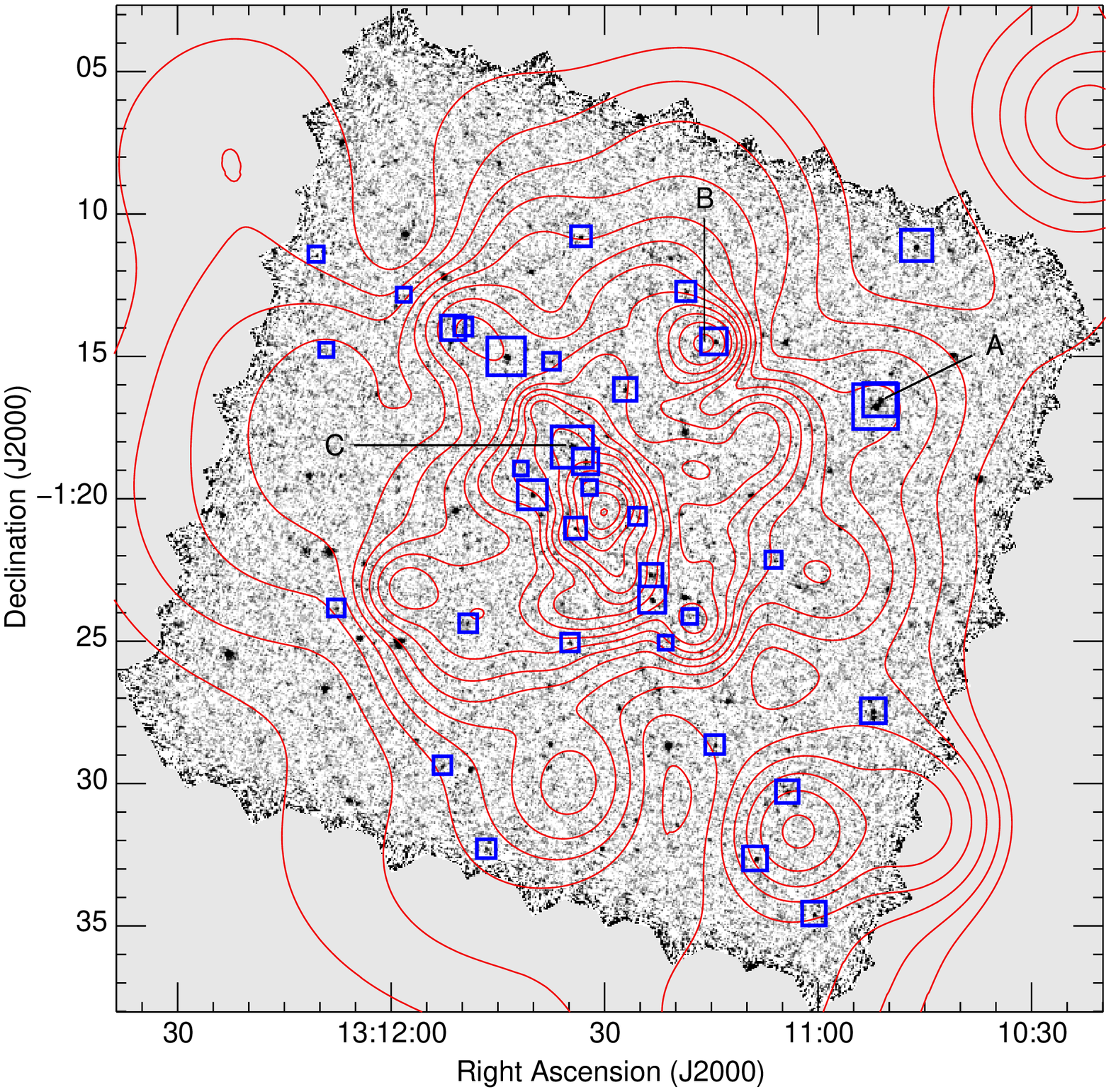}\includegraphics[height=101mm]{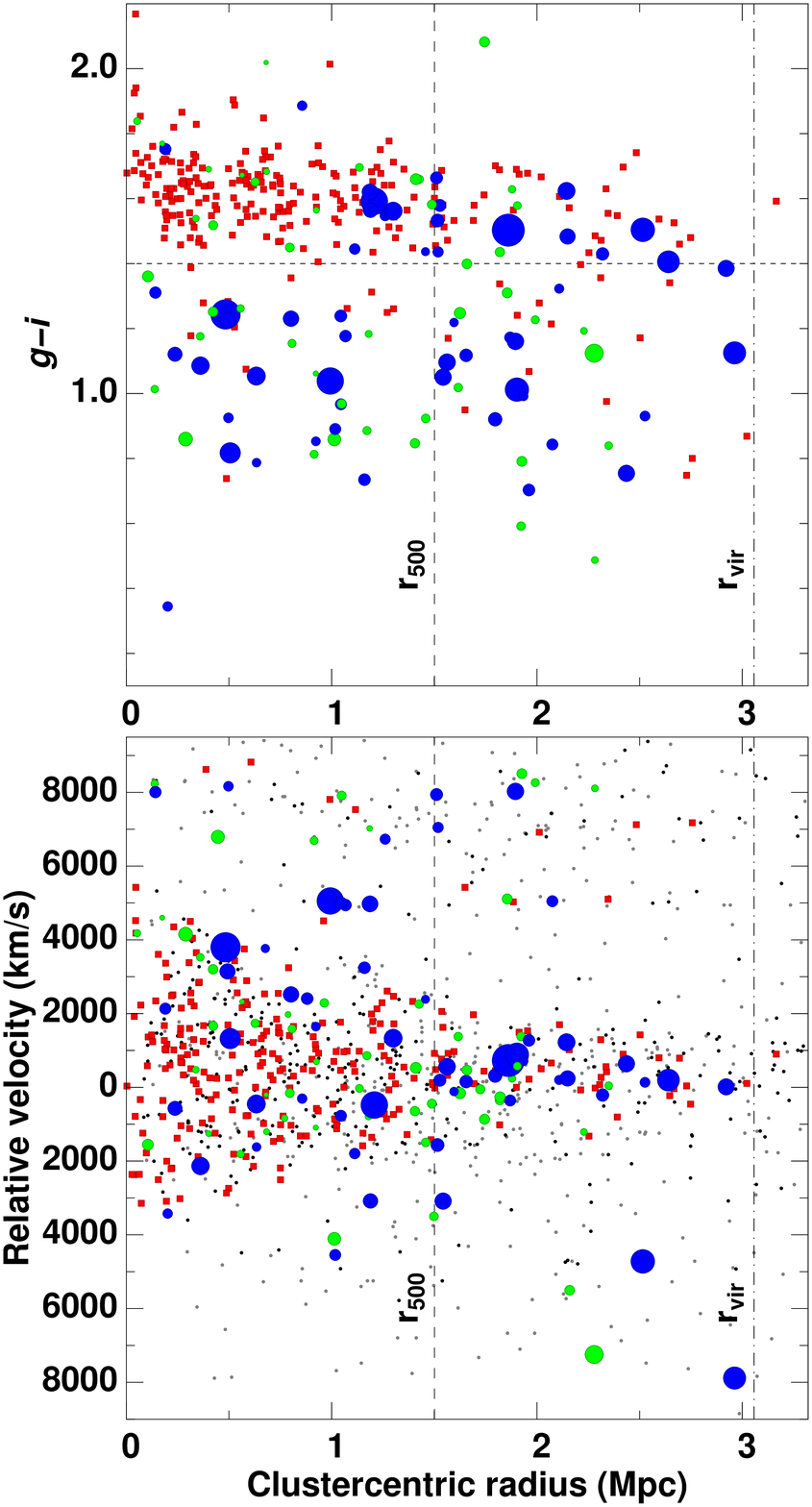}
\caption{ ({\em left:}) The {\em Herschel}/PACS 100$\mu$m map, with
 the spectroscopically confirmed cluster members indicated by blue
 open squares (scaling as 100$\mu$m flux). Red
 contours indicate the projected $K$-band luminosity density
 of confirmed cluster members (weighted to account for
 spectroscopic incompleteness) estimated using the adaptive kernel method. 
 ({\em top-right}) Optical $g{-}i$
  colour versus clustercentric radius. 
  ({\em bottom-right}) Redshift versus projected clustercentric radius.
Blue, green and red symbols 
  respectively indicate $K{\le}K^{*}{+}2$ galaxies detected by: both {\em Herschel} and 
{\em Spitzer}; just {\em Spitzer}; and neither telescope.
  Small
  black and grey points indicate low-mass ($K{>}K^{*}{+}2$) from ours
  and Czoske's spectroscopic surveys.  }
\label{cmradius}
\label{phasespace}
\label{map}
\end{figure*}

In this letter we present an analysis of the {\em Herschel} (Pilbratt
et al. \cite{pilbratt}) FIR imaging of Abell 1689 at $z{=}0.1832$, one
of the first clusters studied in the infrared with {\em ISO} (Fadda et
al. \cite{fadda00}; Balogh et al. \cite{balogh}, hereafter B02; Duc et al. \cite{duc}, hereafter D02).
Abell 1689 has many characteristics of a dynamically relaxed cluster,
with a regular and highly symmetric X-ray morphology (Xue \& Wu
\cite{xue}), the centroid of which coincides both with the brightest cluster 
galaxy and the centre of mass derived from strong lensing (Limousin et
al. \cite{limousin}).  However, the significant discrepancies obtained
between mass estimates derived from strong lensing and X-ray analyses
(Andersson \& Madejski \cite{andersson}), the very high concentration
value observed for the cluster, and its large
($\theta_{E}{\sim}45^{\prime\prime}$) Einstein radius, all strongly
favour a triaxial model in which the cluster is aligned along the
line-of-sight (Oguri et al. \cite{oguri}; Corless, King \& Clowe 
\cite{corless}).  Moreover, a substructure comprising ${\sim}1$5\%
of the total cluster mass (Riemer-S{\o}rensen et al. \cite{riemer})
1.5--2$^{\prime}$ to the NE of the cluster centre has been identified
from dynamical, X-ray and strong lensing analyses (Czoske
\cite{czoske}; Andersson \& Madejski \cite{andersson}; Limousin et
al. \cite{limousin}), while Kawaharada et al. (\cite{kawaharada},
hereafter K10) find evidence for an accreting filament on
10--18\,arcmin scales.

We present the data in \S2, results in \S3, and summary and discussion
in \S4.  
Throughout we assume a cosmology with $\Omega_{M}{=}0.3$,
$\Omega_{\Lambda}{=}0.7$ and H$_{0}{=}70$\,km\,s$^{-1}$\,Mpc$^{-1}$.

\section{Data}

A field of $25^{\prime}{\times}25^{\prime}$ centred on Abell 1689 was
mapped with the Photodetector Array Camera and Spectrometer (PACS;
Poglitsch et al. \cite{poglitsch}) at 100$\mu$m and 160$\mu$m in scan
map mode on 23 December 2009. The images were reduced using {\sc hipe} (Ott
\cite{ott}) and sources detected using {\sc SExtractor} (Bertin et
al. \cite{bertin}) with 100 and 160$\mu$m fluxes estimated using fixed
circular apertures of diameter 12 and 16\,arcsec respectively. The
resulting catalogues are 90\% complete at 18\,mJy at 100$\mu$m and
28\,mJy at 160$\mu$m. For more details see Smith et
al. (\cite{smith10}).

The same $25^{\prime}{\times}25^{\prime}$ field had been 
observed at 24$\mu$m with {\em Spitzer}/MIPS to a sensitivity of
400$\mu$Jy (Haines {\em et al.} \cite{haines09a}). Stellar masses and
optical/NIR absolute magnitudes of cluster galaxies were determined
from Sloan Digital Sky Survey $ugriz$ and UKIRT/WFCAM $JK$-band
photometry.  
A mosaic of 15 {\em Hubble Space Telescope} WFPC2
F616W+F814W images from B02 cover the central $9.8^{\prime}{\times}7.7^{\prime}$ region, from which B02 and D02 obtained morphological classifications of the cluster galaxies.


Over March--May 2009 we obtained 1006 spectra with MMT/Hectospec, targeting probable cluster members (based on their $M_{K}$ and $J{-}K$ colour as in Haines et al. \cite{haines09a}) with $K{<}K^{*}{+}2$ over the full WFCAM field, prioritising those detected at 24$\mu$m, and giving the highest priority to all
sources with $f_{24}{>}1$\,mJy (irrespective of their NIR colour), identifying 387 cluster members. A1689 already had extensive spectroscopic coverage, most notably from Czoske (\cite{czoske}) and the survey of ISOCAM detections by D02, taking us up to a total of 989 unique cluster members. Via this strategy, we have redshifts for 75 of 83 Herschel detections with
$f_{100}{>}20$\,mJy, and 45 of 46 with $f_{100}{>}35$\,mJy (the one 
remaining source appears to be at $z{\ga}1$).

For each {\em Herschel} detection we 
calculate the total infrared flux produced by summing together the {\em Spitzer} 24$\mu$m and the two {\em Herschel} passbands, giving a global FIR measure,
$f_{FIR}{=}f_{24}{\Delta}{\nu}(24\mu m){+}f_{100}{\Delta}{\nu}(100\mu
m){+}f_{160}{\Delta}{\nu}(160\mu m)$. The observed $f_{FIR}$ were
compared to those predicted for each of the 9 luminosity-dependent
infrared SEDs of Rieke et al. (\cite{rieke}) for
$10{<}{\log}(L_{IR}/L_{\odot}){<}12$ placed at the cluster redshift,
and by interpolating between the two nearest SED models, obtain an
estimate for the total infrared luminosity
$L_{IR}$(5--1000$\mu$m). For those sources not detected at 160$\mu$m,
the estimate of $L_{IR}$ was made directly from $f_{100}$. For A1689
our {\em Herschel} completeness limits correspond to
$L_{IR}{=}2.5{\times}10^{10}L_{\odot}$. We note that our incomplete
coverage over 5--80$\mu$m, adjoint with an apparent
100$\mu$m excess seen by {\em Herschel} (Pereira et al. \cite{pereira}; Rawle
et al. \cite{rawle}; Smith et al. \cite{smith10}) in comparison to the model
SEDs of Rieke et al. (\cite{rieke}), could result in our $L_{IR}$ values being
overestimated by 10--20\%.  In total we detect 39 members of Abell
1689 with {\em Herschel}, with 100$\mu$m fluxes and infrared
luminosities in the range 10--156\,mJy and
1.5--2$1{\times}10^{10}L_{\odot}$ respectively.

We look for the AGN contribution to the {\em Herschel} cluster sources
by matching to X-ray point-source catalogues from both XMM data that
covers the entire {\em Herschel} map (K10) and deeper {\em Chandra}
data covering the cluster core to
$L_{X}($0.5--8\,keV$){=}2{\times}10^{41}$erg\,s$^{-1}$ (Gilmour et
al. \cite{gilmour}). Just three X-ray sources are members of A1689,
two of which are radio-loud ellipticals with no detectable IR
emission. The last ($L_{X}{=}5.5{\times}10^{41}$erg\,s$^{-1}$),
identified by D02 as a Seyfert, has $f_{24}{=}0.97$\,mJy, but no
corresponding {\em Herschel} detection. By measuring directly the
pixel values at the galaxy's location, we estimate
$f_{100}{=}7{\pm}4$mJy. We therefore conclude that all of the {\em
  Herschel} sources within A1689 are powered by star formation.

\section{Results}

\begin{figure}
\centering
\includegraphics*[width=65mm]{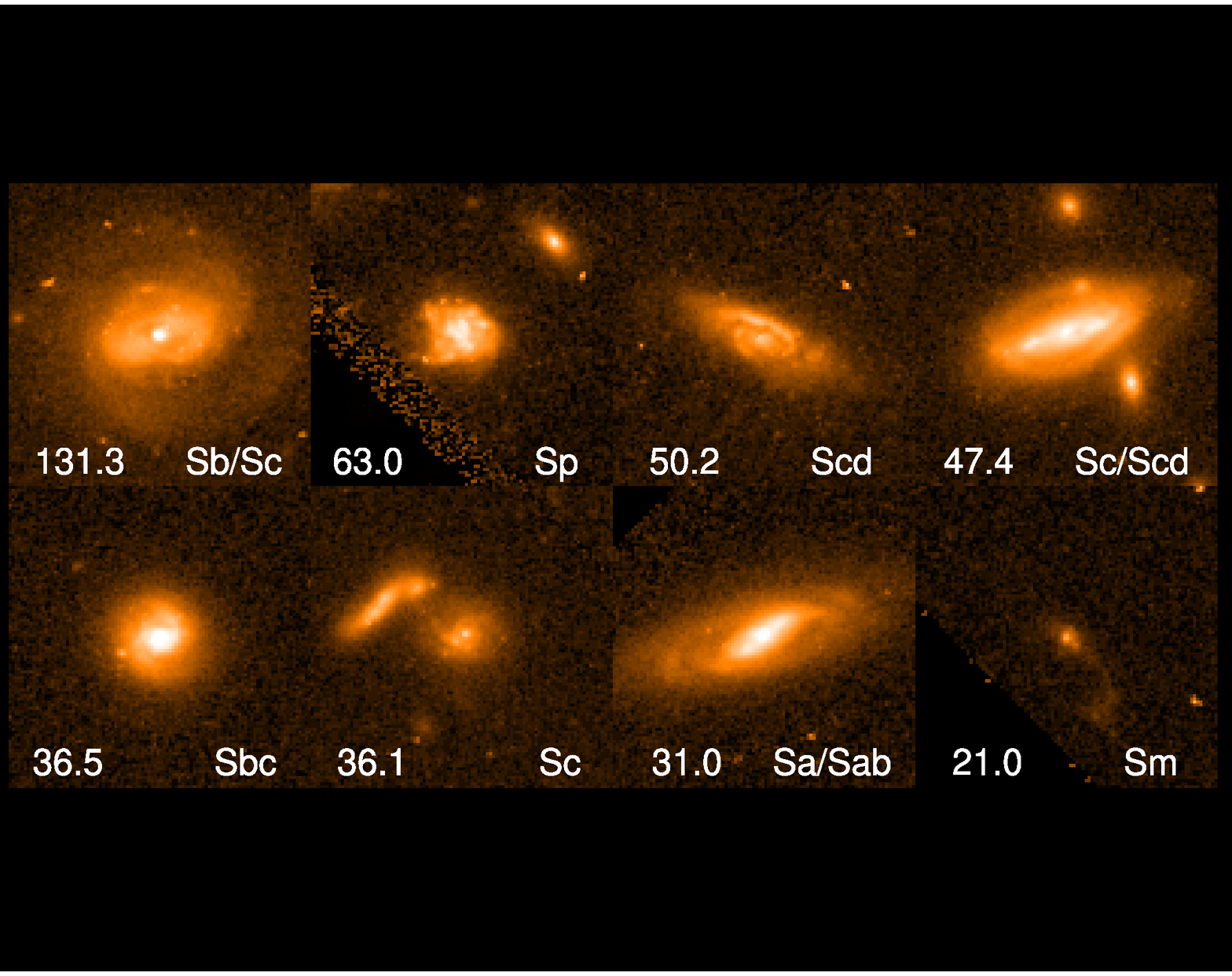}
\caption{{\em HST}/WFPC2 thumbnails of {\em Herschel} sources in A1689, with both 100$\mu$m fluxes (in mJy) and morphologies from B02/D02 indicated.}
\label{morph}
\end{figure}

\subsection{Spatial distribution of star-forming galaxies}

The {\em Herschel}/PACS 100$\mu$m map is presented in Fig.~\ref{map},
with the confirmed cluster members indicated by blue squares. The red
contours instead show the $K$-band light distribution of
The cluster $K$-band light is clearly elongated along the NE/SW direction
on scales 2--10\,arcmin, while a number of galaxy groups in the infall
regions can also be identified. Many of the {\em Herschel} sources are
also aligned along this same axis, forming an apparent filamentary
structure extending over 6\,Mpc across the full 100$\mu$m map from the
NE to the SW corner. Using the 2-dimensional Kolmogorov-Smirnov test 
(Fasano \& Franceschini \cite{fasano}) we find the spatial distribution 
of {\em Herschel} sources consistent with that of $K$-band selected cluster galaxies ($P_{KS}{=}0.83$), but
inconsistent with being uniformly distributed across the 100$\mu$m map ($P_{KS}{=}0.997$) or having a circularly symmetric distribution about 
the cluster centre ($P_{KS}{=}0.96$), confirming the apparent alignment of the {\em Herschel} sources.

B02 found the velocity distribution of the {\em ISO}-detected and
emission-line galaxies in A1689 to be bimodal, avoiding the central
velocity peak, and identified them as infalling.  We thus
investigate the dynamics of the {\em Herschel}-detected galaxies
within A1689 by plotting the line-of-sight relative velocity versus
clustercentric radius in the bottom-right panel of
Fig.~\ref{phasespace}. Galaxies belonging to A1689 are demarcated by
two clean caustics, forming the classic ``trumpet''-shape flaring out
to relative velocities of ${\sim}4$\,000\,km\,s$^{-1}$, and
allowing easy identification of the cluster population. Most of the
{\em Herschel} galaxies lie along or adjacent to these caustics
indicative of an infalling population (Reg\H{o}s \& Geller
\cite{regos}), consistent with B02 and Lemze et al. (\cite{lemze}) who
find the velocity anisotropy profile for A1689 to become predominately
radial at $r\,{\ga}1$\,Mpc. This fits with Biviano \& Katgert
(\cite{biv04}) who show that late-type/emission-line cluster galaxies
are on mostly radial profiles and are absent for $r\,{\la}0.2\,r_{200}$.

\subsection{Morphologies and star formation rates}

\begin{figure}
\centering
\includegraphics[width=65mm]{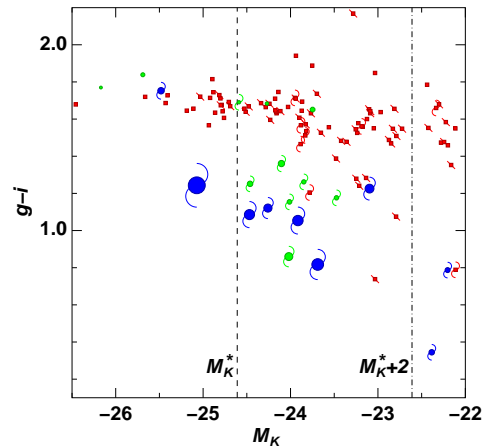}
\caption{Colour-magnitude diagram for A1689 galaxies within the {\em
    HST} mosaic. Symbols are coloured as in Fig.~\ref{cmradius} ({\em right panels}) and have sizes
  scaling with $L_{IR}$ where available.  Plain circles indicate
  ellipticals, barred-symbols indicate S0s, and spirals indicate
  late-types.}
\label{cm}
\end{figure}

Figure~\ref{morph} shows all eight {\em Herschel}-detected cluster galaxies within the {\em HST} mosaic of B02. All are unremarkable $L^{*}$ spirals, barring one
low-mass merger. It is notable that
these are largely late-type spirals (Sbc or later) as opposed to the
early-type spirals (Sab or earlier) identified by Geach et al. (2009)
as LIRGs in Cl\,0024+16 at $z{=}0.4$, and hence it seems unlikely we are
seeing the same populations at differing epochs. In Figure~\ref{cm} we
plot $g{-}i$ versus $M_K$ for the cluster galaxies within the HST
mosaic. The ``red sequence'' is dominated early-types (E/S0s),
but those spirals found on the red sequence are mostly IR-dim,
indicating that their red colour is due to lack of star-formation
rather than dust extinction.  Instead, the vast majority of the
star-forming spirals identified by {\em Herschel} and {\em Spitzer}
lie in the blue cloud, indicating that, at least in the cluster core,
the identification of red sequence and blue cloud populations with
passive and star-forming galaxy populations is largely valid in the
infrared.

We used the star formation rate (SFR) calibration of Calzetti et
al. (\cite{calzetti}), which uses $L(24\mu$m) and $L(H\alpha)$ to
measure the obscured and unobscured components of star formation in a
galaxy.  The H$\alpha$ luminosities of each galaxy were obtained using
Eq.~5 from Hopkins et al. (\cite{hopkins}), which combines the
EW(H$\alpha$) with $M_{r}$ as the measure of the global continuum
flux. We obtain SFRs in the range 1--10\,M$_{\odot}{\rm yr}^{-1}$
(assuming a Kroupa IMF), consistent with the {\em ISO}-based values of
D02.

\begin{figure}
\centering
\includegraphics[width=70mm]{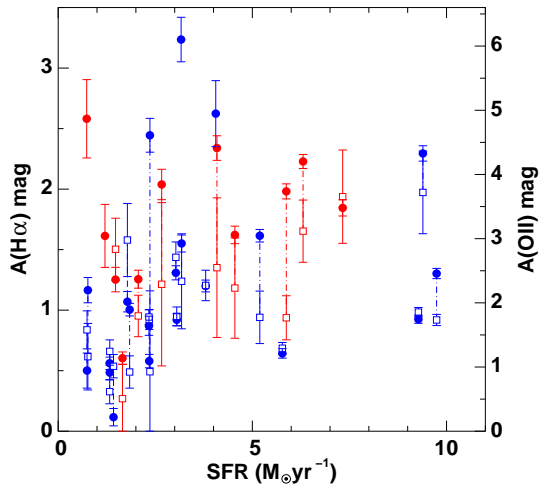}
\caption{SFRs and A(H$\alpha$) for 100$\mu$m-detected cluster members within the red sequence ($g{-}i\,{>}1.4$; red symbols) or blue cloud ($g{-}i\,{<}1.4$; blue).} 
\label{aha}
\end{figure}

In Fig.~\ref{aha} we plot the obtained SFRs and A(H$\alpha$) estimates
for the cluster {\em Herschel} sources, in which the levels of
extinction are derived either from the relative contribution of the
H$\alpha$ and 24$\mu$m emission to the SFR estimator (solid circles),
or the Balmer decrement measured from the optical spectrum (open
squares).  We identify two largely distinct populations of
star-forming galaxies: the blue cloud population ($g{-}i{<}1.4$; blue
symbols) in which the level of extinction is close to the
canonical 1 mag often used to correct H$\alpha$ fluxes; and a dusty
red population in which the star-formation is highly obscured with
A(H$\alpha){\sim}2$\,mag, reddening the galaxy sufficiently for it to
appear on the red sequence ($g{-}i{>}1.4$; red symbols), although in
all cases we detect significant H$\alpha$ emission. For most galaxies
we find the two methods for determining A(H$\alpha$) to be quite
consistent, but we also identify a population of dusty red galaxies
for which the Balmer decrement severely underestimates the level of
obscuration (and hence the SFRs), demonstrating the need for
mid/far-IR data to reveal the true nature of star formation in
all galaxies.

\section{Summary and Discussion}

The results of the {\em Herschel}/PACS photometry can be interpreted to 
suggest a filament of
star-forming galaxies feeding the cluster A1689 at $z{=}0.1832$ from
both sides. The direction of this filamentary structure coincides with
the NE/SW axis of elongation of the cluster $K$-band light, and the
observed asymmetry/substructure seen on 1--2\,arcmin scales in
previous X-ray and strong lensing analyses. K10 found anisotropies in
gas temperature and entropy distributions in the cluster outskirts
(10--18$^{\prime}$), with a high $T_{X}{\sim}5.4$\,keV seen in the NE
direction consistent with accretion flow along a filament, but low gas
temperature (${\sim}1.7$\,keV) in all other directions. They also
found evidence for a connecting large-scale structure (${\sim}2$\,deg)
in the same NE direction in the form of SDSS galaxies with photometric
redshifts consistent with the cluster.  The 100$\mu$m selection has
made the filaments stand out clearly against the background and dense
cluster region, in the same way as seen by Fadda et
al. (\cite{fadda08}) for 24$\mu$m sources in Abell 1763, due to the
enhanced star-forming activity among the filament galaxies.

The SFRs of the {\em Herschel} sources in A1689 are in the range
1--10\,M$_{\odot}{\rm yr}^{-1}$, much lower than the 20--40\,M$_{\odot}{\rm yr}^{-1}$ seen by
Geach et al. (\cite{geach06}, \cite{geach}) for LIRGs in the 
rich cluster
Cl\,0024+16 at $z{=}0.4$. If this star formation is to build the
bulges needed to turn the spirals infalling into clusters at
$z{\sim}0.4$ into the S0s found in local clusters it would require
${\ga}$1--3\,Gyr at the SFRs observed here and in other $z{\sim}0.2$
clusters (Smith et al. \cite{smith10}), rather than the
${\sim}10^{8}$yr implied by Geach et al. (\cite{geach})'s results.

\begin{figure}
\centering
\includegraphics[width=68mm]{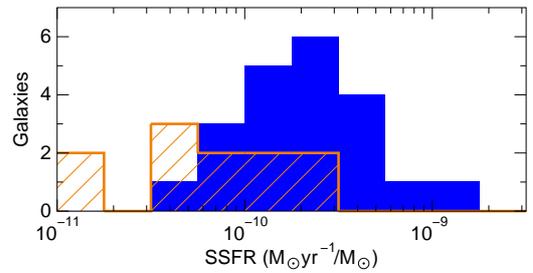}
\caption{Histograms of specific-SFRs for 100$\mu$m-detected cluster members within the red sequence (orange stripes) and blue cloud (blue).}
\label{ssfr}
\end{figure}

We find two populations of {\em Herschel} sources in A1689, analogous
to the blue cloud and dusty-red galaxies identified by Wolf et
al. (\cite{wolf05},\cite{wolf09}) in A901/2, and we find similar
EW(O{\sc ii}), EW(H$\delta$) and specific-SFRs to them.  In
particular, for the dusty red galaxies we obtain $\langle$\,EW(O{\sc
  ii})$\rangle{\sim}5${\AA}, due to having A(O{\sc ii}$){\sim}4$\,mag
placing them on the k+a/e(a) boundary, yet their specific-SFRs are
systematically ${\sim}3{\times}$ lower than those of the blue cloud
population (Fig.~\ref{ssfr}), suggesting that they are currently being
quenched. Whereas the ``blue cloud'' population are found throughout
the cluster, the ``dusty red'' galaxies are found only outside the
cluster core ($r{\ga}1$\,Mpc), where they make up half of the
star-forming galaxies detected by {\em Herschel} (Fig.~\ref{cmradius};
top-right). This is consistent with the findings of STAGES that the
dusty red galaxies prefer the intermediate-density environment of
cluster infall regions (Gallazzi et al. \cite{gallazzi}; Wolf et
al. \cite{wolf09}), and morphologically they found ${\sim}70$\% to be
Sa/Sb spirals with a bright nucleus or inner bar/disk, as well as a
few cases of interactions or merger remnants, suggestive of galaxy
harassment, interactions or mergers for their origin.

Four of the brightest eight {\em Herschel} detections appear
associated with groups: two are within a compact system of 3 galaxies
($\sigma_{\nu}{\sim}120$\,km\,s$^{-1}$; labelled A in Fig.~\ref{map});
one within a $\sigma_{\nu}{\sim}320$\,km\,s$^{-1}$ group (16 members;
group B); and the last within the substructure identified by Czoske
(\cite{czoske}; group C). These, plus similar results obtained by
Pereira et al. (\cite{pereira}) for Abell 1835, suggest that
merger-induced starbursts and transformations in low-mass groups
(pre-processing) could represent a significant contribution to the
high-SFR cluster galaxy population, and an important evolutionary
pathway for the formation of S0s.  The availability of FUV--FIR imaging
plus extensive spectroscopy for a sample of 30 clusters spanning the full range of morphologies and dynamical states within LoCuSS will allow us in future to
estimate the relative fractions of infalling galaxies accreted by
clusters from filaments, groups and the field, and measure the
importance of pre-processing inside groups.

\begin{acknowledgements}
  CPH thanks STFC for support. GPS is supported by the Royal Society.
  Support for MJP and EE was provided by NASA through an award issued
  by JPL/Caltech.
\end{acknowledgements}

\end{document}